\documentclass[pre,epsfig,twocolumn,showpacs,preprintnumbers,amssymb]{revtex4}
\usepackage{bm,graphicx,amsmath}

\begin{document}
\title{Diffusive transport of light in two-dimensional granular materials}

\author{Zeinab Sadjadi$^1$, and MirFaez Miri$^{2}$}
\email{miri@iasbs.ac.ir}
\affiliation{$^{1}$Theoretische Physik, Universit\"{a}t des Saarlandes,
66041 Saarbr\"{u}cken, Germany
\\ $^{2}$Department of Physics, University of Tehran, P.O. Box 14395-547, Tehran, Iran}

\begin{abstract}
We study photon diffusion in a two-dimensional random packing of
monodisperse disks as a {simple} model of granular material. We apply ray optics approximation to set up a persistent random walk for the photons.
We employ Fresnel's intensity reflectance with its rich dependence on the incidence angle and polarization state of the light.
We present an {\it analytic} expression for the transport-mean-free path $l^{\ast}$ in terms of the refractive indices of grains and host medium, grain radius,
and packing fraction. We perform numerical simulations to examine our analytical result.
\end{abstract}


%45.70.-n    granular systems
%81.05.Rm    Porous materials, granular materials
%42.25.Dd    Wave propagation in random media
%05.40.Fb    Random walk

%42.68.Ay    Propagation, transmission, attenuation, and radiative transfer
%82.70.Rr    Aerosols and foams

\maketitle

\section{Introduction}

Pebbles on a sea shore, sand, rice, and sugar, are a few examples of ubiquitous granular systems\ \cite{duran,ristow}.
Granular media consists of discrete particles of size larger than $100 ~ \mathrm{\mu m}$, interacting with each other through dissipative contact forces.
In the absence of an external drive, particles rapidly lose their kinetic energy.
However, granular materials under external forces exhibit transition between a liquid-like and a solid-like state.
In order to understand jamming transition\ \cite{Liu}, size segregation\ \cite{Ottino}, convection rolls, pattern formation and dynamical instabilities\ \cite{78641} of
granular systems, probing the micron-scale dynamics of constituent particles is important. Although granular systems are {\it opaque},
diffusing wave spectroscopy (DWS)\ \cite{DWS} non-invasively
probes their dynamics\ \cite{menon1,menon2,menon3,menon4,swi,cra1,cra2,cra3,cra4,Pak1,Pak2,z1,z2,z3,z4,leutz}.

In a turbid material, light experiences many scattering events before
leaving the sample, and the transport of light energy is
diffusive\ \cite{sheng}. Consequently, the photon can be considered
as a random walker. The structural details of the opaque medium reflect in the transport-mean-free path $l^{\ast}$, over
which the photon direction becomes randomized. Moreover, the dynamics of scatterers leads to the temporal intensity fluctuations in the
speckle field of the multiply scattered light. Utilizing the intensity auto-correlation, DWS determines
$l^{\ast} $ and the mean-squared displacement of the scatterers. DWS has been used to study colloidal dispersions\ \cite{DWS},
liquid crystals\ \cite{DWSnematic}, biopolymers\ \cite{biop}, and foams\ \cite{Durianold,hoo1}.

Multiple light scattering from grains has been invoked to reveal their
relative motion. Three-dimensional gravity-driven granular flow\ \cite{menon1}, gas-fluidized beds\ \cite{menon2}, water-fluidized beds\ \cite{swi},
vibro-fluidized systems\ \cite{menon3,Pak1,Pak2,z1,z2,z3,z4},
avalanche flow\ \cite{menon4}, creeping motion\ \cite{cra1}, dilation due to temperature variations\ \cite{cra3},
and response to a localized compression force\ \cite{cra4} have been thoroughly studied.
As a key parameter, $l^{\ast}$ was measured. For glass beads dispersed in water, $l^{\ast} \approx 14R-16R$
for $80 ~\mathrm{ \mu m} \leqslant R\leqslant 200 ~ \mathrm{ \mu m} $. These samples had a packing fraction $ \phi \approx 0.64 $\ \cite{leutz}.
For glass spheres of radius $R=47.5 ~ \mathrm{ \mu m}$ dispersed in air, $l^{\ast} \approx 15 R $ is reported\ \cite{menon1}.
Crassous\ \cite{cra2} developed a ray-tracing program to access $l^{\ast}$, but only for packing fraction $\phi \approx 0.64$.

It is natural to study $l^{\ast}$ as a
function of the refractive index of grains $n_{in}$, refractive index of the host medium $n_{out}$, average grain radius $R$,
and packing fraction $\phi$. In one approach, expansion of electromagnetic fields in a series of vector multipole fields\ \cite{bor},
or other accurate techniques can be invoked to simulate the speckle pattern.
To extract $l^{\ast}$, simulations must be repeated for a large number of realizations of the random system.
Numerical calculations are quite demanding as each sample contains hundreds of grains.
In another approach, we focus on the elucidation of mechanisms underlying the random walk of photons.
As grains are much larger than the wavelength of light, we rely on ray optics approximation.
Within this framework, the role of {\it reflection} and {\it total internal reflection} phenomena in abrupt change of photons' paths, can be pictured.

We study photon diffusion in a two-dimensional random packing of
monodisperse disks as a {simple} model of granular material.
We assume that the grains are homogenous and transparent. We employ ray optics approximation to follow a light beam or
photon as it is reflected by the grains with a probability called the intensity reflectance.
The photon's random walk based on the above rules is a persistent random walk\ \cite{w1}. This shows that the diffusive transport of light in granular media and
foams\ \cite{miriA,zmrs} are much similar. As an extension of our previous study\ \cite{zmrs}, here we take into account that the intensity reflectance depends on the incidence angle and the polarization state of the light. Writing master equations to describe the photon transport, we
obtain {\it analytic} expression for $l^{\ast}$ as a function of model parameters $n_{in}$, $n_{out}$, $R$, and $\phi$. We perform numerical simulations to examine
our analytical result.

Our article is organized as follows. In Section~\ref{model} we
introduce the model. Photon transport in a random
packing of disks using Fresnel's intensity reflectance is discussed
in Sec.~\ref{mm}. Discussions, conclusions, and an outlook are
presented in Sec.~\ref{dis}.

\section{Model} \label{model}

Following a step-by-step approach to reality, we deliberately focus on {\it two}-dimensional granular systems.
Our model granular medium is a random packing of circular disks. All nonoverlapping disks have the same radius
$R$, and cover a fraction $\phi$ of the plane. The refractive indices of grains and host medium are $n_{in}$ and $n_{out}$, respectively.
We assume that $n_{in} > n_{out}$.

Grains ($>100 ~ \mathrm{\mu m}$) are larger than the visible light wavelength ($400-700 ~ \mathrm{n m}$).
Therefore, we employ ray optics. A light beam or photon experiences transmission or reflection as it hits the surface of a grain.
We denote by $r_{o\rightarrow i}$ the intensity reflectance for photons moving in the host medium and hitting a grain.
Similarly, we denote by $r_{i \rightarrow o}$ the intensity reflectance for photons moving in a gain and hitting its surface.
According to the Fresnel's formulas, both $r_{o\rightarrow i}$ and $r_{i \rightarrow o}$ depend on the polarization state of the light, incidence angle $\gamma$, $n_{out}$ and $n_{in}$, see Appendix\ \ref{appa}.
Indeed the Fresnel's intensity reflectance $r_{i \rightarrow o}(\gamma)$ is $1$ for $\gamma> \gamma_c$, where
\begin{equation}
\gamma_c=\arcsin(n_{out}/n_{in})
\end{equation}
is the critical angle. Our model does respect the total internal reflection phenomena.

It is instructive to consider a toy model, a one-dimensional lattice of (point) grains.
On hitting a grain, a photon will be either reflected by probability $r$ or persist on its direction of motion\ \cite{miriB}.
This leads to a persistent random walk of the photons, where the walker remembers its direction from the previous step\ \cite{w1,w2}.
Now it is clear that implementing the rules of ray optics for photon diffusion in two- and three-dimensional granular media results in
a generalized persistent random walk. Firstly introduced by F\"{u}rth as a model for diffusion in a number of
biological problems\ \cite{furth}, and shortly after by Taylor in the
analysis of turbulent diffusion\ \cite{tay}, the persistent random walks
are employed in polymers\ \cite{flory}, Landauer
diffusion coefficient for a one-dimensional solid\ \cite{55}, and general transport mechanisms\ \cite{t0b}.

\section{Photon transport in a two-dimensional granular material}\label{mm}

\subsection{Analytical treatment} \label{AA}

To simplify our analytical treatment of photons random walk,
we further assume that (i) The length of photon steps outside (inside) the grains is ${L}_{out}$ (${L}_{in}$),
{\it i.e.} we neglect the fact that the length of photon steps are not equal.
(ii) A photon which transmits into or out of a grain, does not change its direction of motion.
In other words, we neglect the fact that the angle of refraction is not equal to the angle of incidence.

\begin{figure}[t]
\includegraphics[width=0.8\columnwidth]{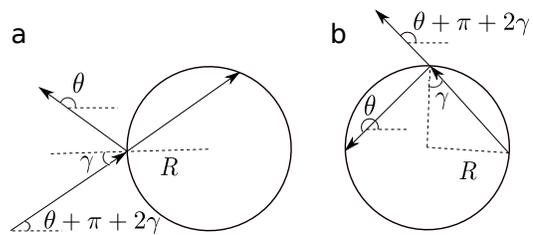}
\caption{(a) Path of a photon moving in the host medium and
hitting a grain with an incidence angle $\gamma$. (b) Path of a
photon moving in a grain and hitting its surface with an incidence
angle $\gamma$. The step length inside the grain is $ 2 R \cos
\gamma$, where $R$ is radius of the grain.
 }
\label{f1}
\end{figure}

In the two-dimensional space, each photon step can be specified by an angle relative to the $x$ axis.
Consider a photon which moves in the host medium along the direction $\theta+\pi+2\gamma$ and hits a grain with an incidence angle $\gamma $, see Fig.~\ref{f1}(a).
The photon will be either reflected to the direction $\theta$ by probability $r_{o\rightarrow i}(\gamma)$, or enter the grain.
The probability distribution of the random variable $ 0 < \gamma<\pi/2 $ is\ \cite{zmrs}
\begin{equation}
F_{out}(\gamma)=cos\gamma.
\end{equation}
Similarly, a photon moving in a grain along
the direction $\theta+\pi+2\gamma$ and hitting its boundary with
an angle $\gamma$, will be either reflected to the direction $\theta$ by probability $r_{i \rightarrow o}(\gamma)$, or leave the grain along the initial direction $\theta+\pi+2\gamma$. Now the probability distribution of the incidence angle is
\begin{eqnarray}
F_{in}(\gamma) &=&\left\{ \begin{array}{lll}
      cos\gamma/sin\gamma_c   & \vert\gamma\vert < \gamma_c\\
         \\
        0 & \mbox{ otherwise,}
     \end{array}
     \right.
\end{eqnarray}
see Appendix\ \ref{appesafe}. Note that $n_{in} > n_{out}$ and $r_{i \rightarrow o}(\gamma)=1$ for $\gamma> \gamma_c$. Thus $\gamma<\gamma_c$ ensures
that a photon moving in the grain is not trapped forever.

We let $P_{n}^{out}(\mathbf{x}|\theta) dx dy$ ($P_{n}^{in}(\mathbf{x}|\theta) dx dy$) denote the probability that a
photon moving along the direction $\theta$, arrives at vicinity $dx dy$ of position
$\mathbf{x}=(x,y)$ after its $n$th step outside (inside) a grain. We express the evolution of
$P_{n}^{out}(\mathbf{x}|\theta)$ and $P_{n}^{in}(\mathbf{x}|\theta)$ by following master equations:
\begin{eqnarray}
P_{n+1}^{out}(\mathbf{x}|\theta)\!\!&=&\!\!\frac{1}{2}\int_{-\frac{\pi}{2}}^{\frac{\pi}{2}}
P_n^{out}( \mathbf{x} \!-\!{L}_{out} \mathbf{e}_\theta |\theta+\pi+2\gamma)  F_{out}(\gamma) \nonumber \\
& &\! \! \times  r_{o\rightarrow i}(\gamma)  d\gamma
+ \overline{t}_{i \rightarrow o} P_n^{in}(  \mathbf{x}\!-\!\!{L}_{out} \mathbf{e}_\theta |\theta) , \label{mastereqa} \\
P_{n+1}^{in}(\mathbf{x}|\theta)\!\! &=&\!\!\frac{1}{2}\int_{-\frac{\pi}{2}}^{\frac{\pi}{2}}
P_n^{in}(\mathbf{x}-{L}_{in} \mathbf{e}_\theta |\theta+\pi+2\gamma)  F_{in}(\gamma) \nonumber \\
& &\!\!  \times   r_{i \rightarrow o}(\gamma)   d\gamma  + \overline{t}_{o\rightarrow i} P_n^{out}(   \mathbf{x}\!-\!\!{L}_{in} \mathbf{e}_\theta |\theta) ,
\label{mastereq}
\end{eqnarray}
where $\mathbf{e}_\theta=(\cos\theta,\sin\theta )$ is the unit vector along the direction $\theta$,
${L}_{out}$ (${L}_{in}$) denotes the average length of photon steps outside (inside) the grains, and
\begin{eqnarray}
&&\overline{t}_{o\rightarrow i}=\frac{1}{2}\int_{-\frac{\pi}{2}}^{\frac{\pi}{2}} \left(1-r_{o\rightarrow i}(\gamma) \right)F_{out}(\gamma)d\gamma , \nonumber\\
&&\overline{t}_{i \rightarrow o}=\frac{1}{2}\int_{-\gamma_c}^{\gamma_c} \left(1-r_{i \rightarrow o}(\gamma) \right)F_{in}(\gamma)d\gamma.
\end{eqnarray}
The first term on the right-hand side of Eq. (\ref{mastereqa}) represents the reflection of
the photon with a probability $r_{o\rightarrow i}(\gamma)$. On arriving at
position $(x-{L}_{out}\cos\theta,y-{L}_{out}\sin\theta )$ along the
direction $\theta+\pi+2\gamma $, the photon changes its direction by an angle
$\pi+2\gamma $ according to the probability distribution $F_{out}(\gamma)$.
The second term focuses on photons which move inside the grain and transmit to the host medium with a
probability $\overline{t}_{i\rightarrow o}$.
The photon which has arrived at position $(x-{L}_{out}\cos\theta,y-{L}_{out}\sin\theta )$ along the direction $\theta$, moves {ballistically} to position $(x,y)$.
Equation (\ref{mastereq}) has a similar interpretation.

We are interested in the second moment of photon distribution with respect to the spatial
coordinates $x$ and $y$. To evaluate the moments of an arbitrary distribution function $P_{n}(x,y|\theta)$,
we utilize its associated characteristic function\ \cite{w1}
\begin{equation}
\mathbf{P}_{n}( \omega_x, \omega_y |m )
 \equiv \int_{-\pi}^{\pi} e^{i m \theta} \int \int e^{i \vec{\omega}
\cdot\mathbf{x}} P_{n}(x,y |\theta ) dx dy  d\theta .\label{charac}
\end{equation}
Indeed
\begin{eqnarray}
\langle x^{k_{1}} y^{k_{2}} \rangle_n &\equiv& \int \int \int
x^{k_{1}} y^{k_{2}}  P_{n}(x,y|\theta) dx dy d\theta \nonumber \\
&=&
 \left. (-i)^{k_{1}+k_{2}} \frac{\partial^{k_{1}+k_{2}}
\mathbf{P}_{n}(\vec{\omega} |m=0 )}{\partial \omega_{x}^{k_{1}}
\partial \omega_{y}^{k_{2}}} \right|_{\vec{\omega}=\mathbf{0}} ,
\label{mean}
\end{eqnarray}
where $ k_1$ and $ k_2$ are either zero or positive
integers, and $\vec{\omega}=(\omega_x,\omega_y)$.

The Fourier transform of master equations (\ref{mastereqa}) and (\ref{mastereq}) are
\begin{eqnarray}
& &{\bf P}_{n+1}^{out}(\omega,\alpha|m) = \nonumber \\
&& \sum_{k=-\infty}^{+\infty}i^k e^{-ik\alpha}J_k(\omega L_{out})  \Big[\overline{t}_{i \rightarrow o} {\bf P}_{n}^{in}(\omega,\alpha|k+m) \nonumber \\
& & ~~~~~~+ c_{m,k}^{out}  {\bf P}_{n}^{out}(\omega,\alpha|k\!+\!m) \Big],  \nonumber \\
&&{\bf P}_{n+1}^{in}(\omega,\alpha|m)=\nonumber \\
& &\sum_{k=-\infty}^{+\infty}i^k e^{-ik\alpha}J_k(\omega L_{in})\Big[\overline{t}_{o\rightarrow i} {\bf P}_{n}^{out}(\omega,\alpha|k+m)  \nonumber \\
& & ~~~~~~ +  c_{m,k}^{in}   {\bf P}_{n}^{in}(\omega,\alpha|k\!+\!m) \Big],
\label{mastereqFourier}
\end{eqnarray}
where  $\omega$ and $\alpha $ are the polar representation of the
vector $\vec{\omega} = (\omega_x, \omega_y)$, $J_k$ is the $k$th-order Bessel function, and
\begin{eqnarray}
c_{m,k}^{out} \!\!&=&\!\!\!\!\frac{(-1)^{m+k}}{2} \!\!\int_{-\frac{\pi}{2}}^{\frac{\pi}{2}}
e^{-i (2m+2k) \gamma}  F_{out}(\gamma)r_{o\rightarrow i}(\gamma)   d\gamma ,\nonumber\\
c_{m,k}^{in} \!\!&=&\!\!\!\!\frac{(-1)^{m+k} }{2}\!\!\int_{-\frac{\pi}{2}}^{\frac{\pi}{2}}
e^{-i (2m+2k) \gamma}  F_{in}(\gamma)  r_{i \rightarrow o}(\gamma)   d\gamma .
\end{eqnarray}
We introduce the Taylor expansions
\begin{eqnarray}
{\bf P}_{n+1}^{out}(\omega,\alpha|m) &\approx& Q_{0,n}^{out}(\alpha|m)+i\omega L_{out}Q_{1,n}^{out}(\alpha|m)  \nonumber\\
&& -\frac{\omega^2 L_{out}^2}{2} Q_{2,n}^{out}+..., \nonumber  \\
{\bf P}_{n+1}^{in}(\omega,\alpha|m) &\approx& Q_{0,n}^{in}(\alpha|m)+i\omega L_{in}Q_{1,n}^{in}(\alpha|m)  \nonumber\\
&&-\frac{\omega^2 L_{in}^2}{2} Q_{2,n}^{in}+...   .
\label{taylor}
\end{eqnarray}
Now we insert Eq. (\ref{taylor}) into Eq. (\ref{mastereqFourier}).
Using the Taylor expansion of the Bessel functions and collecting all terms with the same power in $\omega$, we obtain a complicated set of
recursion relations for $Q_{i,n}^{out}(\alpha|m)$ and $Q_{i,n}^{in}(\alpha|m)$, see Appendix\ \ref{appb}.
There is an elegant method to transform this set of coupled linear difference equations to a set of algebraic equations:
the $z$ transform\ \cite{w1,jury}. The $z$ transform
${Q}(z)$ of a function $Q_n$ of a discrete variable $n=0, 1, 2, ...$ is defined by
\begin{equation}
{Q}(z)=\sum_{n=0}^{\infty} Q_n z^n .
\end{equation}
The $z$ transform of Eqs. (\ref{Q0})-(\ref{Q2}) leads to
a set of algebraic equations whose solutions $Q_i^{in}(z|\alpha,m)$ and $Q_i^{out}(z|\alpha,m)$ are reported in Appendix\ \ref{appb}.

From Eqs. (\ref{mean}) and (\ref{taylor}) it follows that the first and second moments of photon distribution are
\begin{eqnarray}
&&\langle x \rangle_n=L_{out}Q_{1,n}^{out}(0,0)+L_{in}Q_{1,n}^{in}(0,0)  ,\nonumber\\
&&\langle y\rangle_n=L_{out}Q_{1,n}^{out}(\frac{\pi}{2},0)+L_{in}Q_{1,n}^{in}(\frac{\pi}{2},0)  , \nonumber\\
&&\langle x^2\rangle_n=L_{out}^2Q_{2,n}^{out}(0,0)+L_{in}^2Q_{2,n}^{in}(0,0),  \nonumber\\
&&\langle y^2\rangle_n=L_{out}^2Q_{2,n}^{out}(\frac{\pi}{2},0)+L_{in}^2Q_{2,n}^{in}(\frac{\pi}{2},0).
\label{averages}
\end{eqnarray}
Powered by the analytical expressions for the inverse $z$ transform of $Q_i^{in}(z|\alpha,m)$ and $Q_i^{out}(z|\alpha,m)$, we find that in the limit
$ n \rightarrow \infty $
\begin{eqnarray}
&&\langle x \rangle_{n}= \langle y \rangle_{n} = 0 \label{mean1}, \nonumber \\
&&\langle x^2\rangle_n =  \langle y^2\rangle_n=( w_{out} L_{out}^2 +w_{in} L_{in}^2  ) n ,\label{x2nb}
\end{eqnarray}
where
\begin{eqnarray}
&&w_{out}=\frac{\overline{t}_{o\rightarrow i}}{\overline{t}_{o\rightarrow i}+\overline{t}_{i \rightarrow o}}  \Big[\frac{\overline{t}_{i \rightarrow o}}{2}\nonumber\\
&&+\frac{  L_{in}L_{out}^{-1} c_{1,-1}^{in}\overline{t}_{i \rightarrow o}+\overline{t}_{i \rightarrow o}\big(  \overline{t}_{i \rightarrow o} \overline{t}_{o\rightarrow i}+c_{0,1}^{out}(1-c_{0,1}^{in})\big)}{(1-c_{1,0}^{out})(1-c_{1,0}^{in})-\overline{t}_{o\rightarrow i}\overline{t}_{i\rightarrow o}}  \Big]\nonumber\\
&&+ \Big[ \frac{ L_{in}L_{out}^{-1} \overline{t}_{o\rightarrow i}\overline{t}_{i \rightarrow o}+c_{1,-1}^{out}\big(  \overline{t}_{i \rightarrow o} \overline{t}_{o\rightarrow i}+c_{0,1}^{out}(1-c_{0,1}^{in})\big) }{(1-c_{1,0}^{out})(1-c_{1,0}^{in})-\overline{t}_{o\rightarrow i}\overline{t}_{i\rightarrow o}}\nonumber\\
&&~~~~ +\frac{c_{0,0}^{out}}{2} \Big] \frac{\overline{t}_{i \rightarrow o}  }{\overline{t}_{o\rightarrow i}+\overline{t}_{i \rightarrow o}},
\end{eqnarray}
\begin{eqnarray}
&&w_{in}=\frac{\overline{t}_{o\rightarrow i}}{\overline{t}_{o\rightarrow i}+\overline{t}_{i \rightarrow o}}  \Big[\frac{c_{0,0}^{in}}{2} \nonumber\\
&&+\frac{  L_{in}^{-1} L_{out} \overline{t}_{o\rightarrow i}\overline{t}_{i \rightarrow o}+c_{1,-1}^{in}\big(  \overline{t}_{i \rightarrow o} \overline{t}_{o\rightarrow i}+c_{0,1}^{in}(1-c_{0,1}^{out})\big) }{(1-c_{1,0}^{out})(1-c_{1,0}^{in})-\overline{t}_{o\rightarrow i}\overline{t}_{i\rightarrow o}}\Big] \nonumber\\
&&+ \Big[\frac{  L_{in}^{-1} L_{out} c_{1,-1}^{out}\overline{t}_{o\rightarrow i}+\overline{t}_{o\rightarrow i}\big(  \overline{t}_{i \rightarrow o} \overline{t}_{o\rightarrow i}+c_{0,1}^{in}(1-c_{1,0}^{out})\big) }{(1-c_{1,0}^{out})(1-c_{1,0}^{in})-\overline{t}_{o\rightarrow i}\overline{t}_{i\rightarrow o}} \nonumber\\
&&~~~~ +\frac{\overline{t}_{o\rightarrow i}}{2} \Big] \frac{\overline{t}_{i \rightarrow o}  }{\overline{t}_{o\rightarrow i}+\overline{t}_{i \rightarrow o}}.
\end{eqnarray}

The task is now expressing the time $n \tau$ spent for the $n$ steps of the random walker. Indeed
\begin{equation}
\tau= f_{out}\tau_{out} +f_{in}\tau_{in},
\end{equation}
where $f_{out}$ ($f_{in}=1- f_{out}$) is the fraction of time that the photons spend outside (inside) the grains,
$\tau_{out}={n_{out}L_{out}}/{c}$ ($\tau_{in}={n_{in}L_{in}}/{c} $) is the average time spent to make a step outside (inside) the grains, and
$c$ is the velocity of light in vacuum.
To evaluate $f_{in}$, we consider a photon hitting a grain with an incidence angle $\gamma$. The probability of $m$ internal steps before leaving the grain is
$ t_{i \rightarrow o}(\gamma) [r_{i \rightarrow o}(\gamma)]^{m-1}$. Thus the photon spends a time
$$\sum_{m=1} m\tau_{in} t_{i \rightarrow o}(\gamma) [r_{i \rightarrow o}(\gamma)]^{m-1}=\tau_{in} /t_{i \rightarrow o}(\gamma)$$ inside a grain before leaving it. Averaging
with respect to the probability distribution $F_{in}(\gamma)$, we find that a photon spends a time
\begin{equation}
\frac{1}{2}\int_{-\gamma_c}^{\gamma_c} \frac{\tau_{in}F_{in}(\gamma) }{t_{i \rightarrow o}(\gamma)}d\gamma=\tau_{in}\langle \frac{1}{t_{i \rightarrow o}}\rangle
\end{equation}
inside the grain. It follows that
\begin{eqnarray}
f_{out } &=&\frac{\tau_{out} }{\tau_{in}\langle \frac{1}{t_{i \rightarrow o}}\rangle+\tau_{out}}  ,\nonumber \\
f_{in} &=&\frac{\tau_{in} \langle \frac{1}{t_{i \rightarrow o}}\rangle}{\tau_{in}\langle \frac{1}{t_{i \rightarrow o}}\rangle+\tau_{out}} .
\label{Alpha}
\end{eqnarray}
We also note that $\phi={{L}_{in}}/({{L}_{in}+{L}_{out}})$. Figure\ \ref{f1}(b) suggests that
${L}_{in}=  \langle 2R \cos\gamma \rangle $ where $\gamma$ is the incidence angle of photons moving in the
grain, and here $\langle...\rangle$ denotes averaging with respect to the
probability distribution $F_{in}(\gamma)$. Hence we
find
\begin{eqnarray}
L_{out} &=& R (\cos \gamma_c + \gamma_c/\sin\gamma_c)  \frac{1-\phi}{\phi}  , \nonumber\\
L_{in} &=&  R (\cos \gamma_c + \gamma_c/\sin\gamma_c). \label{tool}
\end{eqnarray}

As already mentioned, in the long-time limit $n \tau \rightarrow \infty$, the behavior of the mean-square displacements is purely
diffusive, i.e.,
\begin{eqnarray}
\langle x^2\rangle_n =\langle y^2\rangle_n = 2 D\tau n.
\end{eqnarray}
In two-dimensional systems, the transport-mean-free path is defined via
\begin{eqnarray}
l^*=2D/v_m,
\end{eqnarray}
where ${v}_m$ is the transport velocity of light in the medium. To a first approximation
\begin{equation}
{v}_m=  (1-\phi)\frac{c}{n_{out}} +\phi \frac{c}{n_{in}} ,
\label{lkj2}
\end{equation}
where $c/n_{out}$ ($c/n_{in}$) is the velocity of light in the host medium (grains) which
cover a fraction $1-\phi$ ($\phi$) of the plane. Now we utilize Eqs. (\ref{x2nb})-(\ref{lkj2}) to derive the diffusion constant and transport-mean-free path in the two-dimensional granular material
\begin{eqnarray}
D \!&=&\! \frac{1}{2} R c \Big( n_{in}/n_{out} \arcsin(n_{out}/n_{in}) \!+\! \sqrt{1 \!-\!(n_{out}/n_{in})^2} \Big)  \nonumber\\
&& \times \Big(  w_{out} (\frac{1\!-\! \phi}{\phi})^2  +w_{in}  \Big) \frac{n_{out}(\frac{1-\phi}{\phi})+n_{in} \langle \frac{1}{t_{i \rightarrow o}}\rangle}{n_{out}^2(\frac{1 -\phi}{\phi})+n_{in}^2 \langle \frac{1}{t_{i \rightarrow o}}\rangle}, \label{zxc}
\end{eqnarray}
\begin{eqnarray}
l^* &=&  R   \Big( n_{in}/n_{out} \arcsin(n_{out}/n_{in})+\sqrt{1 \!-\!(n_{out}/n_{in})^2} \Big)      \nonumber\\
&& \times \Big(  w_{out} (\frac{1-\phi}{\phi})^2  +w_{in}  \Big) \frac{n_{out}(\frac{1-\phi}{\phi})+n_{in} \langle \frac{1}{t_{i \rightarrow o}}\rangle}{n_{out}^2(\frac{1-\phi}{\phi})+n_{in}^2 \langle \frac{1}{t_{i \rightarrow o}}\rangle} \nonumber\\
&& \times  \frac{1}{\frac{\phi}{n_{in}} + \frac{(1-\phi)}{n_{out}}}.
\end{eqnarray}
We emphasize that $w_{out}$, $w_{in}$ and $\langle \frac{1}{t_{i \rightarrow o}}\rangle$ can be explicitly expressed in terms of
physical parameters $n_{out}$, $n_{in}$, and $\phi$.
The incident electric field can be either perpendicular $\bot$ or parallel $\|$ to the two-dimensional plane covered by the grains.
The intensity reflectance depends on the polarization state of light, thus the diffusion constants $D_\bot$ and $D_\|$ are not equal.

\subsection{Numerical simulations} \label{BB}
We carry out numerical simulations to examine our analytical result for the diffusion constant $D$.
Using contact dynamics simulations\ \cite{Shaebani08jcp},
we generate homogeneous and monodisperse random packing of disks.
We study samples with packing fraction $ \phi \in [0.15, 0.25,...,0.65]$. Each sample consists of $10^4$ nonoverlapping disks.
We let $10^4$ photons perform a random walk in each sample. We launch the photons in a direction specified by angle $ \theta_0$.
We repeat the simulation for all angles $ \theta_0\in [30^{\circ}, 60^{\circ}, ..., 360^{\circ}]$.
We implement Fresnel's formulas and Snell's law as a photon hits a grain surface.
Following a standard Monte Carlo procedure, we generate the trajectory of each photon, and evaluate the statistics of the photon cloud at different times to access the diffusion constant $D$. For improving
the speed of our ray tracing program, we adopt the cell index
method commonly used in the molecular dynamics simulations, see Refs.\ \cite{miriA} and\ \cite{zmrs} for details.

\begin{figure}
\includegraphics[width=1.0\columnwidth]{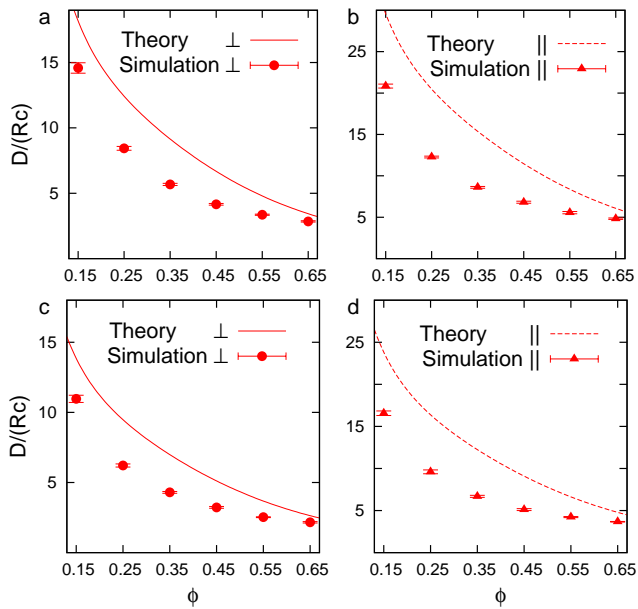}
\caption{(Color online) The diffusion constant (a) $D_\bot$, (b) $D_\|$ (in units of the disk
radius $R$ times the velocity of light $c$) as a function of the
packing fraction $\phi$ for the case $n_{in}=1.5$ and $n_{out}=1.0$.
(c) $D_\bot$ and (d) $D_\|$ for the case $n_{in}=2.0$ and $n_{out}=1.34$. }
\label{f2}
\end{figure}

As an example, we consider the glass disks
($n_{in}=1.5$) immersed in the air ($n_{out}=1.0$).
Figures~\ref{f2}(a) and~\ref{f2}(b) demonstrate $D_\bot$ and $D_\|$ as a function of $\phi$, respectively.
As another example, we consider glass disks ($n_{in}=2$) immersed in the water ($n_{out}=1.34$).
Corresponding $D_\bot$ and $D_\|$ as a function of $\phi$ are shown in Figs.~\ref{f2}(c) and~\ref{f2}(d), respectively.
Equation~(\ref{zxc}) involves no free parameters, but reasonably agrees with the numerical results.

\section{Discussion}\label{dis}

We have studied diffusive light transport in a two-dimensional packing of monodisperse disks. We employed ray
optics to follow a light beam or photon as it is reflected by the disks. We used Fresnel's intensity reflectance with its rich dependence on the incidence angle and polarization state of the light. We note that a photon which moves in a grain, hits its surface with an incidence angle $\gamma$ less than $\gamma_c=\arcsin(n_{out}/n_{in})$. Indeed $\gamma<\gamma_c$ ensures that photons are not caged in grains due to the total internal reflection phenomena.
Using a {\it constant} intensity reflectance independent of the incidence angle\ \cite{zmrs}, it is not clear how the total internal reflection phenomena influences the transport-mean-free path $l^{\ast}$. Moreover, a {constant} intensity reflectance model does not take into account that
$r_{i \rightarrow o}(\gamma)  \neq r_{o \rightarrow i}(\gamma)$.

Our analytical estimate of $D$ is bigger than our numerical estimate by a factor about $1.5$, see Fig.~\ref{f2}.
In writing master equations for photon transport, we do not consider the inequality of step lengths either in the host medium or in the grains.
We also neglect the fact that the angle of refraction is not equal to the angle of incidence.
But in our numerical simulations, the photon step length exhibits its natural distribution. Moreover, we strictly obey Snell's law.

Consider a photon moving in the host medium and hitting a grain with an incidence angle $\gamma$. The photon experiences an average scattering angle
$\int_0^{\pi/2}(\pi-2\gamma)F_{out}(\gamma)d\gamma=2$ (in radians) due to the reflection.
Taking into account the Snell's law, the average angle between the incident and refracted ray is
$\int_0^{\pi/2}[\gamma-\arcsin(\frac{n_{out}}{n_{in}}sin\gamma)]F_{out}(\gamma)d\gamma=0.22 $, where have assumed $n_{in}=1.5$ and $n_{out}=1.0$. Similarly, a photon moving in a grain experiences an average scattering angle $\int_0^{\gamma_c}(\pi-2\gamma)F_{in}(\gamma)d\gamma=2.45$ due to reflection,
and an average scattering angle $\int_0^{\gamma_c}[\arcsin(\frac{n_{in}}{n_{out}}sin\gamma)-\gamma]F_{in}(\gamma)d\gamma=0.22 $ due to transmission.
Thus, the transmission is less efficient than the reflection in randomizing the direction of photons.
Apparently, respecting path randomization due to transmissions, our theoretical $D$ decreases towards numerical one.
We have also studied the distribution function of step length $G({L}_{out})$.
After reaching its pronounced maximum, $G({L}_{out})$ decays exponentially, see Fig. 5 of Ref.\ \cite{zmrs}. Focusing on a radically different model, an uncorrelated random walk in a dilute packing of point scatterers,
Heiderich {\it et al.}~\cite{PLA} found that a broader distribution of step lengths leads to a greater $D$.
The interplay between the two above mentioned opposing impacts on $D$ remains to be clarified.

\begin{figure}
\includegraphics[width=0.7\columnwidth]{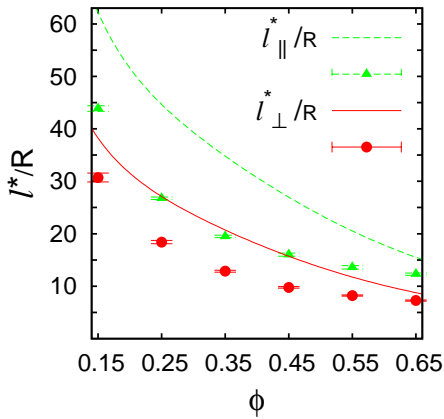}
\caption{ (Color online) The transport-mean-free paths $l^{\ast}_\bot$ and $l^{\ast}_\|$ (in units of the disk
radius $R$) as a function of the
packing fraction $\phi$ for the case $n_{in}=1.5$ and $n_{out}=1.0$.
Theoretical and Monte Carlo simulation results are denoted, respectively, by lines and points.}
\label{f3}
\end{figure}

In two-dimensional space $D= l^{\ast} {v}_m /2$, where ${v}_m$ is the transport velocity
of light. In a medium composed of spheres comparable to
the light wavelength, the transport
velocity differs by an order of magnitude from the phase velocity\
\cite{speed}. The difference between the two velocities is
unimportant when spheres are much larger than the light
wavelength. Equation (\ref{lkj2}) presents a "mean field" estimate of ${v}_m$.
Figure~\ref{f3} demonstrates the transport-mean-free paths $l^{\ast}_\bot=2 D_\bot /{v}_m $ and $l^{\ast}_\|=2 D_\| /{v}_m $ as a function of the
packing fraction $\phi$ for the case $n_{in}=1.5$ and $n_{out}=1.0$. As expected, $l^{\ast}_\bot$ and $l^{\ast}_\|$ monotonically decrease as $\phi$ increases:
The photon hits more gains and rapidly forgets its initial direction of motion.
For packing fraction $\phi \approx 0.64$, we find a reasonable value $l^{\ast}\simeq (l^{\ast}_\bot +l^{\ast}_\|)/2 \simeq 12 R$, cf. Refs.\ \cite{menon1,leutz,cra2}.
Note that to address photon diffusion in a {three}-dimensional system, we have taken an average over $\bot$ and $\|$ polarizations.

To achieve a better understanding of photon diffusion in granular systems, there is still much to do.
We aim at a superior model, which considers path randomization due to transmissions and the natural distribution of step lengths.
Also an extension to the three-dimensional packing of polydisperse spheres is
envisaged. Here we estimate the transport-mean-free path. DWS experiments also reveal the dynamics of granular systems.
Recent theories concerning deformations of a sphere packing, need an improvement as $n_{out}/n_{in}$ becomes less than $2/3$ and the critical angle $\gamma_c=\arcsin(n_{out}/n_{in})$ deviates more from $\pi/2$, see Fig. 5 of Ref.\ \cite{cra2}. Here one must consider
correlation between paths of a photon which moves in a disk and suffers many reflections\ \cite{cra2}.
Currently, we are studying the abrupt change of photon paths due to the local rearrangement of grains. Note that in another system, foam,
bubble rearrangements cause fluctuations in the intensity of scattered light\ \cite{Durianold}.

Different experiments may be performed on {\it two}-dimensional granular systems.
One can immerse a packing of disks in various liquids to study $l^{\ast}$ dependence on the refractive index $n_{out}$.
The incident electric field can be either perpendicular or parallel to the two-dimensional plane.
A ray maintains its polarization state on hitting a disk. Quite remarkably $l^{\ast}_\| > l^{\ast}_\bot$, see Fig.~\ref{f3}.
This is reasonable, since $r_ \| < r_\bot$: A photon sooner forgets its initial direction of motion as the scattering events becomes more probable.
In another anisotropic system, nematic liquid crystal,
the light diffuses faster along the director than perpendicular to the director\ \cite{DWSnematic}.

\appendix
\section{Intensity reflectances}\label{appa}
Fresnel's intensity reflectance depends on
the polarization state of the light. The incident electric field can be either perpendicular $\bot $ or parallel $\|$ to the two-dimensional plane covered by the grains.
For all angles $ 0 < \gamma<\pi/2 $,
\begin{eqnarray}
&&r_{o\rightarrow i}^{\bot}(\gamma)=
\Big\vert\frac{n_{out}cos\gamma-n_{in}\sqrt{1-(\frac{n_{out}}{n_{in}}sin\gamma)^2}}{n_{out}cos\gamma+n_{in}\sqrt{1-(\frac{n_{out}}{n_{in}}sin\gamma)^2}}\Big\vert^2,\nonumber\\
&&r_{o\rightarrow i}^{\|}(\gamma)=
\Big\vert\frac{n_{out}\sqrt{1-(\frac{n_{out}}{n_{in}}sin\gamma)^2}-n_{in}cos\gamma}{n_{out}\sqrt{1-(\frac{n_{out}}{n_{in}}sin\gamma)^2}+n_{in}cos\gamma}\Big\vert^2.\nonumber\\
\end{eqnarray}
For all angles $\gamma<\gamma_c$,
\begin{eqnarray}
&&r_{i\rightarrow o}^{\bot}(\gamma)=
\Big\vert\frac{n_{in}cos\gamma-n_{out}\sqrt{1-(\frac{n_{in}}{n_{out}}sin\gamma)^2}}{n_{in}cos\gamma+n_{out}\sqrt{1-(\frac{n_{in}}{n_{out}}sin\gamma)^2}}\Big\vert^2,\nonumber\\
&&r_{i\rightarrow o}^{\|}(\gamma)=
\Big\vert\frac{n_{in}\sqrt{1-(\frac{n_{in}}{n_{out}}sin\gamma)^2}-n_{out}cos\gamma}{n_{in}
\sqrt{1-(\frac{n_{in}}{n_{out}}sin\gamma)^2}+n_{out}cos\gamma}\Big\vert^2.\nonumber\\
\end{eqnarray}
Indeed $r_{i\rightarrow o}^{\bot}(\gamma) = r_{i\rightarrow o}^{\|}(\gamma)=1$ if $\gamma>\gamma_c$.

\section{The probability distribution $F_{in}(\gamma)$}\label{appesafe}

A photon which moves in a grain hits its surface with an angle $\gamma<\gamma_c$, thus $F_{in}(\gamma)=0$ if $\gamma>\gamma_c$.
To find the probability distribution $F_{in}(\gamma)$ for $\gamma<\gamma_c$, we consider path of photons inside the disk, see
Fig.~\ref{f1}(b). Each ray can be characterized by its distance
$s$ from the center of the disk. $s=R \sin\gamma$ thus $ s < R \sin\gamma_c$.
We assume that the random
variable $s$ has a uniform distribution in the interval $[0, R \sin\gamma_c]$.
The cumulative distribution function $F_c(\gamma)\equiv \int_{0}^{\gamma} F_{in}(\psi) d\psi $ is then $
F_c(\gamma)=\text{Prob}(s < R \sin\gamma)=R \sin \gamma/(R \sin\gamma_c)$. It
follows that
\begin{equation}
F_{in}(\gamma)=\frac{d F_c(\gamma)}{d \gamma}=\cos \gamma/\sin\gamma_c \label{A1}
\end{equation}
for $\gamma<\gamma_c$. Further numerical simulations confirm our
analytical results for $F_{in}(\gamma)$ and $F_{out}(\gamma)$.

\section{Functions $Q_{i,n}^{out}$ and $Q_{i,n}^{in}$}\label{appb}
The functions $Q_{i,n}^{out}$ and $Q_{i,n}^{in}$ introduced in Eq. (\ref{taylor}), are the solutions of the following
equations:

\begin{eqnarray}
& & Q_{0,n+1}^{out}(\alpha|m) = \overline{t}_{i \rightarrow o}Q_{0,n}^{in}(\alpha|m) + c_{m,0}^{out}Q_{0,n}^{out}(\alpha|m),\nonumber\\
& & Q_{0,n+1}^{in}(\alpha|m)=	\overline{t}_{o\rightarrow i}Q_{0,n}^{out}(\alpha|m) + c_{m,0}^{in}Q_{0,n}^{in}(\alpha|m), \nonumber\\
\label{Q0}
\end{eqnarray}

\begin{eqnarray}
& &Q_{1,n+1}^{out}(\alpha|m)= \nonumber\\
&&\lambda\overline{t}_{i \rightarrow o}Q_{1,n}^{in}(\alpha|m) \!+\! c_{m,0}^{out}Q_{1,n}^{out}(\alpha|m) \nonumber\\
&&+\frac{1}{2}e^{-i\alpha}\Big( \overline{t}_{i \rightarrow o} Q_{0,n}^{in}(\alpha|m+1) \!+\!  c_{m,1}^{out}Q_{0,n}^{out}(\alpha|m+1)\Big) \nonumber\\
&&+\frac{1}{2}e^{i\alpha}\Big( \overline{t}_{i \rightarrow o} Q_{0,n}^{in}(\alpha|m-1) \!+\!  c_{m,-1}^{out}Q_{0,n}^{out}(\alpha|m-1)\Big), \nonumber\\
&& Q_{1,n+1}^{in}(\alpha|m) = \nonumber\\
&&\lambda^{-1}\overline{t}_{o\rightarrow i}Q_{1,n}^{out}(\alpha|m) \!+\! c_{m,0}^{in}Q_{1,n}^{in}(\alpha|m) \nonumber\\
&&+\frac{1}{2}e^{-i\alpha}\Big( \overline{t}_{o\rightarrow i} Q_{0,n}^{out}(\alpha|m+1) \!+\!  c_{m,1}^{in}Q_{0,n}^{in}(\alpha|m+1)\Big) \nonumber\\
&&+\frac{1}{2}e^{i\alpha}\Big( \overline{t}_{o\rightarrow i} Q_{0,n}^{in}(\alpha|m-1) \!+\!  c_{m,-1}^{in}Q_{0,n}^{in}(\alpha|m-1)\Big), \nonumber\\
\label{Q1}
\end{eqnarray}

\begin{eqnarray}
& & Q_{2,n+1}^{out}(\alpha|m)= \nonumber\\
&&\lambda^2 \overline{t}_{i \rightarrow o}Q_{2,n}^{in}(\alpha|m)+ c_{m,0}^{out}Q_{2,n}^{out}(\alpha|m) \nonumber\\
&&+e^{-i\alpha}\Big(\! \lambda\overline{t}_{i \rightarrow o} Q_{1,n}^{in}(\alpha|m+1)  \!+\!  c_{m,1}^{out}Q_{1,n}^{out}(\alpha|m+1)\Big) \nonumber\\
&&+e^{i\alpha}\Big(\! \lambda\overline{t}_{i \rightarrow o} Q_{1,n}^{in}(\alpha|m-1)  \!+\!  c_{m,-1}^{out}Q_{1,n}^{out}(\alpha|m-1)\Big) \nonumber\\
&&+\frac{1}{2}\Big(\overline{t}_{i \rightarrow o}Q_{0,n}^{in}(\alpha|m) \!+\! c_{m,0}^{out}Q_{0,n}^{out}(\alpha|m)\Big)\nonumber \\
&&+\frac{1}{4}e^{-2i\alpha}\Big( \overline{t}_{i \rightarrow o}Q_{0,n}^{in}(\alpha|m+2) \!+\! c_{m,2}^{out}Q_{0,n}^{out}(\alpha|m+2) \Big)\nonumber\\
&&+\frac{1}{4}e^{2i\alpha}\Big( \overline{t}_{i \rightarrow o}Q_{0,n}^{in}(\alpha|m-2) \!+\! c_{m,-2}^{out}Q_{0,n}^{out}(\alpha|m-2) \Big),\nonumber\\
&&Q_{2,n+1}^{in}(\alpha|m)=\nonumber\\
&&\lambda^{-2} \overline{t}_{o\rightarrow i}Q_{2,n}^{out}(\alpha|m)+ c_{m,0}^{in}Q_{2,n}^{in}(\alpha|m) \nonumber\\
&&+e^{-i\alpha}\Big(\! \lambda^{-1}\overline{t}_{o\rightarrow i} Q_{1,n}^{out}(\alpha|m+1) \!+\!  c_{m,1}^{in}Q_{1,n}^{in}(\alpha|m+1)\Big) \nonumber\\
&&+e^{i\alpha}\Big(\! \lambda^{-1}\overline{t}_{o\rightarrow i} Q_{1,n}^{out}(\alpha|m-1) \!+\!  c_{m,-1}^{in}Q_{1,n}^{in}(\alpha|m-1)\Big) \nonumber\\
&&+\frac{1}{2}\Big(\overline{t}_{o\rightarrow i}Q_{0,n}^{out}(\alpha|m) \!+\! c_{m,0}^{in}Q_{0,n}^{in}(\alpha|m)\Big)\nonumber \\
&&+\frac{1}{4}e^{-2i\alpha}\Big( \overline{t}_{o\rightarrow i}Q_{0,n}^{in}(\alpha|m+2) \!+\! c_{m,2}^{in}Q_{0,n}^{in}(\alpha|m+2) \Big)\nonumber\\
&&+\frac{1}{4}e^{2i\alpha}\Big( \overline{t}_{o\rightarrow i}Q_{0,n}^{out}(\alpha|m-2) \!+\! c_{m,-2}^{in}Q_{0,n}^{in}(\alpha|m-2) \Big),\nonumber\\	
\label{Q2}
\end{eqnarray}
where $\lambda={L_{in}}/{L_{out}}$. The above set of linear difference equations express $Q_{i,n+1}^{out}$ and $Q_{i,n+1}^{in}$ in terms of $Q_{i,n}^{out}$ and $Q_{i,n}^{in}$
The $z$ transform of $Q_{n+1}$ is simply $Q(z)/z-Q_{n=0}/z$. Note the similarities of this rule with the Laplace
transform of the time derivative of a continuous function\ \cite{arf}. The $z$-transform of above difference equations leads to
a set of {\it algebraic} equations whose solutions are
\begin{eqnarray}
&&Q_0^{out}(\!z|\alpha,\!m\!) \!\!=\!\!\frac{(1\!\!-\!\!z c_{m,0}^{in})Q_{0,n=0}^{out}(\alpha,\!m)\!\!+\!\!z\overline{t}_{i \rightarrow o}Q_{0,n=0}^{in}(\alpha,\!m\!) }{\Delta_m},\nonumber\\
&&Q_0^{in}(\!z|\alpha,\!m\!) \!\!=\!\!\frac{z\overline{t}_{o\rightarrow i}Q_{0,n=0}^{out}(\alpha,\!m\!)\!+\!(1\!\!-\!\!z c_{m,0}^{out})Q_{0,n=0}^{in}(\alpha,\!m)\!\! }{\Delta_m},\nonumber\\
\label{Q0z}
\end{eqnarray}

\begin{eqnarray}
&&Q_1^{out}(z|\alpha,\!m\!)\!\!=\frac{(1\!\!-\!\!z c_{m,0}^{in})A_m(\alpha,z)+\lambda z\overline{t}_{i \rightarrow o}B_m(\alpha,z) }{\Delta_m},\nonumber\\
&&Q_1^{in}(z|\alpha,\!m\!)\!\!=\frac{\lambda^{-1}z\overline{t}_{o\rightarrow i}A_m(\alpha,z)+(1\!\!-\!\!z c_{m,0}^{out})B_m(\alpha,z) }{\Delta_m},\nonumber\\
\label{Q1z}
\end{eqnarray}

\begin{eqnarray}
&&Q_2^{out}(z|\alpha,\!m\!)\!\!=\frac{(1\!\!-\!\!z c_{m,0}^{in})D_m(\alpha,z)\!+\!\lambda^2 z\overline{t}_{i \rightarrow o}E_m(\alpha,z) }{\Delta_m},\nonumber\\
&&Q_2^{in}(z|\alpha,\!m\!)\!\!=\frac{\lambda^{-2}z\overline{t}_{o\rightarrow i}D_m(\alpha,z)+(1\!\!-\!\!z c_{m,0}^{out})E_m(\alpha,z) }{\Delta_m}.\nonumber\\
\label{Q2z}
\end{eqnarray}
Here
\begin{eqnarray}
&&A_m(\alpha,z)=  Q_{1,n=0}^{out}(\alpha,m) \nonumber\\
&&+\frac{z}{2}e^{-i\alpha}\Big( \overline{t}_{i\rightarrow o} Q_{0,n}^{in}(\alpha|m+1)+  c_{m,1}^{out}Q_{0,n}^{out}(\alpha|m+1)\Big) \nonumber\\
&&+\frac{z}{2}e^{i\alpha}\Big( \overline{t}_{i\rightarrow o} Q_{0,n}^{in}(\alpha|m-1)+  c_{m,-1}^{out}Q_{0,n}^{out}(\alpha|m-1)\Big) ,\nonumber
\end{eqnarray}

\begin{eqnarray}
&&B_m(\alpha,z)= Q_{1,n=0}^{in}(\alpha,m)\nonumber\\
&&+\frac{z}{2}e^{-i\alpha}\Big( \overline{t}_{o\rightarrow i} Q_{0,n}^{in}(\alpha|m+1)+  c_{m,1}^{in}Q_{0,n}^{in}(\alpha|m+1)\Big) \nonumber\\
&&+\frac{z}{2}e^{i\alpha}\Big( \overline{t}_{o\rightarrow i} Q_{0,n}^{out}(\alpha|m-1)+  c_{m,-1}^{in}Q_{0,n}^{in}(\alpha|m-1)\Big),\nonumber
\end{eqnarray}

\begin{eqnarray}
&&D_m(\alpha,z)= Q_{2,n=0}^{out}(\alpha,m)\nonumber\\
&&+ ze^{-i\alpha}\Big( \lambda\overline{t}_{i\rightarrow o} Q_{1,n}^{in}(\alpha|m+1)+  c_{m,1}^{out}Q_{1,n}^{out}(\alpha|m+1)\Big) \nonumber\\
&&+ze^{i\alpha}\Big( \lambda\overline{t}_{i\rightarrow o} Q_{1,n}^{in}(\alpha|m-1)+  c_{m,-1}^{out}Q_{1,n}^{out}(\alpha|m-1)\Big) \nonumber\\
&&+\frac{z}{2}\Big(\overline{t}_{i\rightarrow o}Q_{0,n}^{in}(\alpha|m) + c_{m,0}^{out}Q_{0,n}^{out}(\alpha|m)\Big)\nonumber \\
&&+\frac{z}{4}e^{-2i\alpha}\Big( \overline{t}_{i\rightarrow o}Q_{0,n}^{in}(\alpha|m+2) + c_{m,2}^{out}Q_{0,n}^{out}(\alpha|m+2) \Big)\nonumber\\
&&+\frac{z}{4}e^{2i\alpha}\Big( \overline{t}_{i\rightarrow o}Q_{0,n}^{in}(\alpha|m-2) + c_{m,-2}^{out}Q_{0,n}^{out}(\alpha|m-2) \Big),\nonumber
\end{eqnarray}

\begin{eqnarray}
&&E_m(\alpha,z)=Q_{2,n=0}^{in}(\alpha,m)\nonumber\\
&&+ze^{-i\alpha}\Big(\! \lambda^{-1}\overline{t}_{o\rightarrow i} Q_{1,n}^{out}(\alpha|m+1) \!\!+\!  c_{m,1}^{in}Q_{1,n}^{in}(\alpha|m+1) \!\Big) \nonumber\\
&&+ze^{i\alpha}\Big( \! \lambda^{-1}\overline{t}_{o\rightarrow i} Q_{1,n}^{out}(\alpha|m-1) \! \!+\!  c_{m,-1}^{in}Q_{1,n}^{in}(\alpha|m-1)\! \Big) \nonumber\\
&&+\frac{z}{2}\Big(\overline{t}_{o\rightarrow i}Q_{0,n}^{out}(\alpha|m) + c_{m,0}^{in}Q_{0,n}^{in}(\alpha|m)\Big)\nonumber \\
&&+\frac{z}{4}e^{-2i\alpha}\Big( \overline{t}_{o\rightarrow i}Q_{0,n}^{in}(\alpha|m+2) + c_{m,2}^{in}Q_{0,n}^{in}(\alpha|m+2) \Big)\nonumber\\
&&+\frac{z}{4}e^{2i\alpha}\Big( \overline{t}_{o\rightarrow i}Q_{0,n}^{out}(\alpha|m-2) + c_{m,-2}^{in}Q_{0,n}^{in}(\alpha|m-2) \Big), \nonumber	
\label{ABDE}
\end{eqnarray}
and
\begin{eqnarray}
\Delta_m=(1-zc_{m,0}^{out})(1-zc_{m,0}^{in})-z^2\overline{t}_{o\rightarrow i}\overline{t}_{i\rightarrow o}.\label{Detm}
\end{eqnarray}

The expressions of $Q_i^{in}(z|\alpha,m)$ and $Q_i^{out}(z|\alpha,m)$ contain the sum of
several terms whose inverse $z$-transform are readily accessible:
\begin{eqnarray}
1 & \leftrightarrow & \frac{1}{1-z} ,\nonumber \\
n  & \leftrightarrow & \frac{z}{(1-z)^2}, \nonumber \\
a^n & \leftrightarrow & \frac{1}{1-a z} , \nonumber \\
n a^n & \leftrightarrow & \frac{az}{(1-a z)^2}.
\end{eqnarray}
Here $a$ is an arbitrary real number whose absolute magnitude is
less than $1$.

\end{document}